Influence of solar eclipse of November 3rd, 2013 on the total ozone column over Badajoz, Spain

D. Mateos*, M. Antón, and J. M. Vaquero

Department of Physics, University of Extremadura, Badajoz, Spain

**\* Corresponding author:** D. Mateos, Departamento de Física, Universidad de Extremadura, Avda. Elvas s/n, 06010, Badajoz, Spain. Telephone: 0034-924289300 (ext. 89122). E-mail: davidmv@unex.es

**Abstract**

The hybrid eclipse of November 3rd, 2013 was observed as partial with a magnitude equal to 0.126 from Badajoz (38º 53' N, 6º 58' W). The evolution of the Total Ozone Column (TOC) values during 4 hours was monitored using a Solar Light Microtops-II manual sun-photometer. Before the eclipse, TOC remained invariable ~280 Dobson Units (DU) for one hour and a half. Once the eclipse was started, a clear decrease in TOC occurred. After the eclipse maximum (with TOC = 273 DU), a rapid TOC recovery was observed. When the eclipse was over, TOC came back to values ~280 DU.

# 1. Introduction

A solar eclipse is always an interesting event calling the attention of scientific community but also general public. The spectacular phenomena of the solar disk obscured by the Moon have captivated the entire humanity for centuries. The impact of the eclipse on the atmosphere is referred to its thermal, chemical and dynamical natures. The change in the incoming radiation achieving the top of the atmosphere modifies the radiative equilibrium and also affects the photochemistry. The solar ultraviolet (UV) radiation is the main natural responsible for the formation and destruction of ozone in the atmosphere. Hence, a change in the radiation available for photochemistry should produce a change in the total ozone column (TOC) and its vertical distribution. However, previous ground-based measurements during solar eclipses showed large discrepancies. For instance, Bojkov (1966) reported an increase in total ozone concentration using a Dobson spectrometer, while Brewer measurements exhibited a substantial total ozone reductions (Zerefos et al., 2000; Kazadzis et al.,2007). Measurements of a NILU multiband instrument also showed an increased in the ozone level (Antón et al., 2010). The reason behind this discrepancy lies on experimental artifact due to the higher contribution of the diffuse irradiance against the decrease in the direct irradiance during the eclipse, being this effect stronger at shorter wavelengths (UV range) (e.g., Koepke et al., 2001).

Taking advantage of this opportunity, the eclipse on November 3rd, 2013, the Department of Physics of the University of Extremadura organized a public event in one of the gardens of "La Alcazaba" (an ancient Moorish citadel) in Badajoz (Spain). A high number of professional and amateur people were involved in the Sun's observation. This event was a big success with almost two hundred people interesting in the solar eclipse. At the same time, instantaneous measurements of TOC were performed by a hand-held

Microtops- II sun-photometer. To our knowledge, this is the first time that this instrument is used in the measure of the total ozone content during a solar eclipse. Although the manual kind of the measurements makes difficult to have a large dataset, a strong effort was made in this event to obtain as much continuous data series as possible.

## 2. Instrumentation

To monitor the evolution of total ozone during the eclipse, the Solar Light Microtops-II manual sun-photometer was used. The Microtops-II measures spectral direct radiation in five channels using five optical collimators of 2.5º field of view: 305.5, 312.5, 320, 936, and 1020 nm (a nominal full width at half maximum of 2.4±0.4 nm). The three channels falling in the UV range are used to retrieve the TOC. For this study, the retrievals obtained with the pair 312.5 and 320 nm were selected to monitor TOC values. The hand-held measurements were performed with the Microtops-II mounted on a tripod. A complete description about the Microtops-II instrument, including its algorithms and way of operating, was given by Morys et al. (2001).

To test the Microtops-II ozone retrievals, a campaign with the cloud-free days during the 25 days previous to the eclipse was performed. Hourly data were measured whenever cloudy conditions allowed the measure. Each individual Microtops-II measurement was averaged among five consecutive scans. Only data showing great stability were taking into account in this study. In this sense, we only considered TOC measurements with a standard deviation less than 5 Dobson Units (DU) in the ozone retrieval. Note that all the points considered in this analysis exhibited standard deviations below 1 DU. These experimental data were validated against satellite-derived

TOC from the Ozone Monitoring Instrument (OMI) (Levelt et al., 2006). OMI overpass data was previously tested in the same region showing small relative differences below 2% with respect to the Brewer instruments (e.g., Antón et al., 2009). The OMI overpass data are free-downloaded by the corresponding website (http://avdc.gsfc.nasa.gov/). Hence, we assumed these TOC data as the reference ones.

Figure 1 shows the evolution of TOC values observed by Microtops-II and OMI instrument over Badajoz site during eight days. From this figure, it seems that the angular response of the Microtops-II seems weakly contribute to the daily averages. Table 1 shows the main statistical parameters derived from the comparison between daily Microtops-II TOC values and OMI data calculated using the methods described by Willmot (1982). The good agreement between Microtops-II and OMI is corroborated with a root-mean-square-error of 3%, and an index of agreement higher than 0.7.

Table 1. Comparison between daily TOC values of Microtops-II and OMI satellite instrument. Statistical parameters calculated following Willmott (1982): $n$ is the number of data, *mbe* in the mean bias error, *rmse* is the root-mean-square-error, $i_{gree}$ is the index of agreement, and *b* and *a* are the slope and intercept of a linear fit.

| Variable | Value |
|---|---|
| $n$ | 8 |
| mbe | -3 % |
| rmse | 3 % |
| $i_{agree}$ | 0.76 |
| b | 0.98 |
| a | -3 DU |

Previous comparisons also showed a good characterization of TOC by the Microtops-II instrument. For instance, Gómez-Amo et al. (2012) at several campaigns in the European Union obtained relative differences in TOC values between Microtops-II and Brewer spectroradiometers smaller than 2%. Earlier comparisons between Microtops-II and OMI data obtained relative differences below 2% in different world regions (e.g., Silva and Tomaz, 2012; Gómez-Amo et al., 2013; Mateos et al., 2014). Hence, this small and manual instrument provides a reliable measure of the total ozone column.

## 3. Eclipse details

The solar eclipse of November 3rd, 2013 belongs to the rare family of hybrid eclipses (or annular/total eclipses). In this kind of eclipses, some sections of the path are annular while other parts are total, with the apparent sizes of the Moon and the Sun very similar varying during the phenomenon.

The path of the hybrid eclipse of 2013 started in the North Atlantic and crossed equatorial Africa, including Gabon, Congo, Democratic Republic of Congo, Uganda, Kenya, and southern Ethiopia. Finally, the lunar shadow leaved the surface of Earth in Somalia. A partial eclipse was visible in eastern North America, northern South America, southern Europe, the Middle East and Africa (see Figure 2).

We observed this eclipse at Badajoz site (38º 53' N, 6º 58' W, red square in Figure 2) located close to the northern limit of the visibility zone. The partial phase of the eclipse began at 11:44 UT and ended at 13:13 UT in Badajoz station. At this location, the partial phase of the eclipse reached a maximum magnitude of 0.126 at 12:28 UT. At this moment, the obscuration of Sun's disc was of 5.3% (maximum solar coverage).

Additional details about the astronomical data of this eclipse were predicted by Espenak and Meeus (2009).

We have computed the observable angular diameter and apparent topocentric coordinates (corrected for refraction) of the Moon and the Sun using the highly accurate ephemerides provided by JPL Horizons On-Line Ephemeris System (http://ssd.jpl.nasa.gov/?horizons). Moreover, the phase f (fraction of the solar disc diameter covered by lunar disc) and the eclipse obscuration g (determined as a covered-to-total solar disc surface ratio) were computed every minute using the equations provided by Szałowski (2002).

## 4. Results and discussion

One of the great advantages of the use of the Microtops-II instrument is the instantaneous characterization of the total ozone. Hence, before-during-after the eclipse event, data were acquired in the public event performed in La Alcazaba and were averaged at 2-minute scale. The total ozone observations started at 10:30h UT and finished at 14:00h UT. Unfortunately, partially cloudy conditions (specially, high clouds) occurred through the solar eclipse. However, the fast acquisition of the Microtops-II instrument allowed a high number of scans (almost 300) to be recorded. Additionally, an exhaustive quality control was established, checking the variability of the aerosol optical depth at 1020 nm derived from the Microtops-II. Since Microtops-II retrieves the aerosol data using direct radiation measurements, high clouds can be misinterpreted as aerosols (e.g., Guerrero-Rascado et al., 2013). Therefore, we filtered those data with AOD at 1020nm over 0.25. Figure 3 shows the evolution of the TOC values during 4 hours on the November 3rd, 2013 and the evolution of the solar eclipse. Before the eclipse, the TOC values remained invariable around 280 DU for one hour and a half. Once the eclipse was started, a clear decrease in the TOC values occurred achieving a minimum value very close to the maximum of the eclipse. The minimum TOC value (~273 DU) is almost a 3% smaller than the value before the eclipse. Hence, in spite of this low solar obscuration, there is a clear TOC decrease. Larger relative values of this decrease, around 10%, were reported for previous total solar eclipses (e.g., Mims and Mims, 1993; Osherovich et al, 1974). For an annular eclipse (80% of solar obscuration), Antón et al. (2010) obtained a relative decrease of TOC around 5% with a Brewer instrument at El Arenosillo (Spain). As was proved by previous studies (e.g., Köepke et al., 2001), the limb darkening effect is wavelength-dependent. Hence, the physical interpretation of the evolution of total ozone can be explained by the change caused by this spectral limb darkening effect, together with the enhance in the diffuse radiance considered as direct component due to the field of view of the Microtops-II during the eclipse (e.g., Kazadzis et al., 2006; Antón et al., 2010).

After the eclipse maximum magnitude, a rapid recovery of the TOC values was observed. At 14:00 h UT, when the eclipse was over, the values of the TOC came back to the earlier eclipse ones (around 280 DU). OMI overpass data for that day was 293.11 DU at 13:19h UT. There was a difference larger than 10 DU than could be attributed to the eclipse itself by the differences in the field of view of this event. Cloudy conditions can also contribute to this discrepancy. Aerosol optical depth seemed to play a minor role that day since the measurements under really cloud-free conditions showed an aerosol load at 1020 nm less than 0.1. Substantial discrepancies between satellite retrievals and ground-based observations are expected for larger AOD (e.g., Mateos et al., 2012).

## 5. Summary and conclusions

The hybrid eclipse of November 3rd, 2013 was observed as partial (maximum magnitude of 0.126) at Badajoz city (Spain). Ground-based total ozone data by Microtops-II manual sun-photometer were selected to monitor TOC evolution during the eclipse. Microtops-II retrievals were validated against OMI satellite data showing a root-mean-square-error of 3% and a high index of agreement. TOC data were recorded during 4 hours. Before the eclipse, TOC was ~280 Dobson Units (DU) for one hour and a half. A decrease in TOC values was observed with the beginning of the solar eclipse, achieving a minimum value of TOC equal to 273 DU at the eclipse maximum. Beyond this point, a rapid TOC recovery was observed and after the eclipse, TOC came back to values ~280 DU.

**Acknowledgements**

Support from the Junta de Extremadura (Research Group Grant No. GR10131) and Ministerio de Economía y Competitividad of the Spanish Government (AYA2011-25945 and CGL2011-29921-C02-01) is gratefully acknowledged. The personal contribution of all the local organization members is also acknowledged: A.J.P. Aparicio, M.L. Cancillo, M.I. Fernández-Fernández, M.C. Gallego, J. A. García, G. Sáez, and A. Serrano. The authors want to thank J.M. Vilaplana (Spanish Institute for Aerospace Technology, INTA) for the lending of Microtops-II instrument (serial number #4710). Manuel Antón thanks 'Ministerio de Ciencia e Innovación' and 'Fondo Social Europeo' for the award of a postdoctoral grant (Ramón y Cajal).

Figure captions

Figure 1. Evolution of TOC values measured by the Microtops-II (circles) and OMI (triangles) instruments.

Figure 2. Map of the hybrid solar eclipse of November 3rd, 2013. Red square indicates the geographical location of Badajoz site.

Figure 3. Evolution of TOC (circles, vertical bars are the standard deviation of each point) recorded by the Microtops-II instrument during the eclipse of November 3rd, 2013, and the obscuration (g, dashed line) and phase (f, solid line) of the solar eclipse. Vertical gray dashed line highlights the maximum magnitude of the eclipse.

Figure + Figure captions

Figure 1

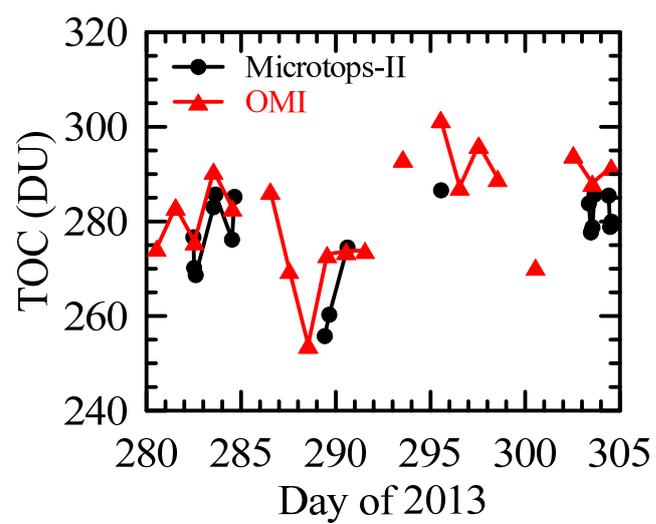

Figure 1. Evolution of TOC values measured by the Microtops-II (circles) and OMI (triangles) instruments.

Figure 2

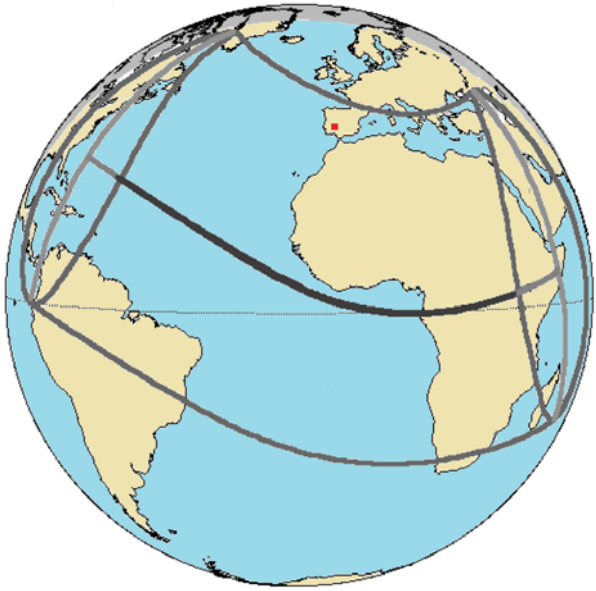

Figure 2. Map of the hybrid solar eclipse of November 3rd, 2013. Red square indicates the geographical location of Badajoz site.

Figure 3

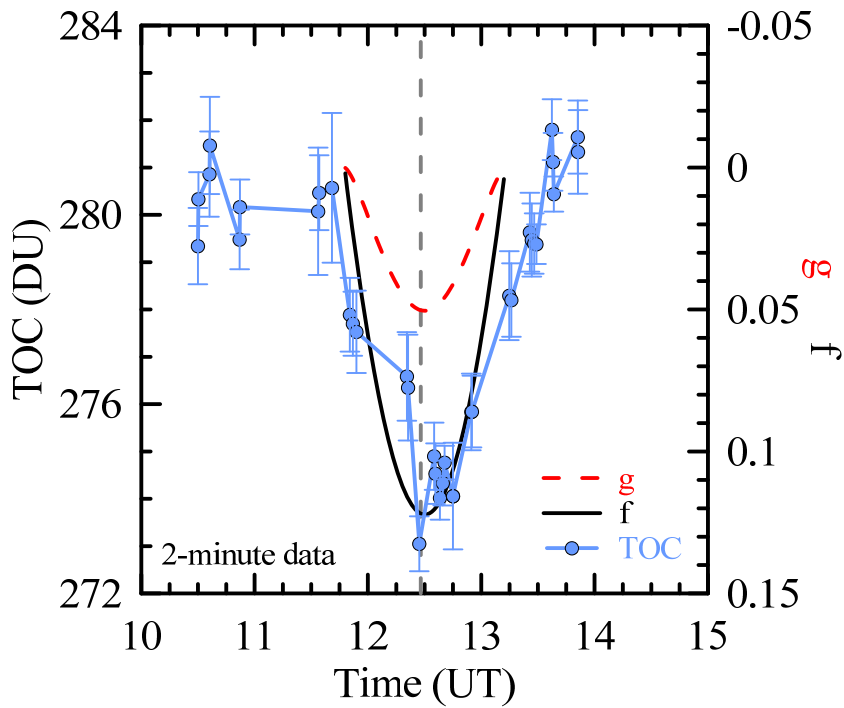

Figure 3. Evolution of TOC (circles, vertical bars are the standard deviation of each point) recorded by the Microtops-II instrument during the eclipse of November 3rd, 2013, and the obscuration (g, dashed line) and phase (f, solid line) of the solar eclipse. Vertical gray dashed line highlights the maximum magnitude of the eclipse.